\documentclass[]{revtex4}
\usepackage{graphicx}
\usepackage{epsfig} 
\usepackage{wrapfig}
\usepackage{amsmath}

\begin{document}
\title{Gamma-ray bursts and collisionless shocks\footnote{Invited review presented at the 33rd annual European Physical Society Conference, Rome 2006.}}

\author{E. Waxman}
\address{Physics Faculty, Weizmann Inst. of Science, Rehovot 76100, Israel}

\begin{abstract}

Particle acceleration in collisionless shocks is believed to be responsible for the production of cosmic-rays over a wide range of energies, from few GeV to $>10^{20}$~eV, as well as for the non-thermal emission of radiation from a wide variety of high energy astrophysical sources. A theory of collisionless shocks based on first principles does not, however, exist. Observations of $\gamma$-ray burst (GRB) "afterglows" provide a unique opportunity for diagnosing the physics of relativistic collisionless shocks. Most GRBs are believed to be associated with explosions of massive stars. Their "afterglows," delayed low energy emission following the prompt burst of $\gamma$-rays, are well accounted for by a model in which afterglow radiation is due to synchrotron emission of electrons accelerated in relativistic collisionless shock waves driven by the explosion into the surrounding plasma. Within the framework of this model, some striking characteristics of collisionless relativistic shocks are implied. These include the generation of downstream magnetic fields with energy density exceeding that of the upstream field by $\sim8$ orders of magnitude, the survival of this strong field at distances $\sim10^{10}$ skin-depths downstream of the shock, and the acceleration of particles to a power-law energy spectrum, $d\log n/d\log\varepsilon\approx-2$, possibly extending to $10^{20}$~eV. I review in this talk the phenomenological considerations, based on which these characteristics are inferred, and the challenges posed to our current models of particle acceleration and magnetic field generation in collisionless shocks. Some recent theoretical results derived based on the assumption of a self-similar shock structure are briefly discussed. 

\end{abstract}
\maketitle

\section{Introduction}
\label{sec:intro}

Due to the low densities characteristic of a wide range of astrophysical environments, shocks observed in many astrophysical systems are collisionless, i.e. mediated by collective plasma instabilities rather than by particle-particle collisions. For example, collisionless shocks play an important role in supernova remnants \citep[e.g.][]{Blandford87}, jets of radio galaxies \citep[e.g.][]{Begelman94,Maraschi03}, gamma-ray bursts \citep[GRBs, e.g.][]{WrevSN,Zhang04,Piran05}, and the formation of the large scale structure of the Universe \citep[e.g.][]{Loeb00,Gruzino01,Schlick03}. Although collisionless shocks have been studied for several decades, theoretically and experimentally, in space and in the laboratory, a self-consistent theory of collisionless shocks based on first principles has not yet emerged \citep[see, e.g., comments in][]{Krall97}.

Observations of GRB "afterglows," the delayed low energy emission following the prompt $\gamma$-ray emission, provide a unique probe of the physics of collisionless shocks. Current understanding suggests that the afterglow radiation observed is the synchrotron emission of energetic, non-thermal electrons in the downstream of a strong collisionless shock driven into the surrounding interstellar medium (ISM) or stellar wind. These collisionless shocks start out highly relativistic, with shock Lorentz factor $\Gamma\sim 100$ on time scale of minutes after the GRB, and gradually decelerate to $\Gamma\sim10$ on a day time scale and $\Gamma\sim1$ on a month time scale. This allows one to probe the physics of the shocks over a wide range of Lorentz factors. 

In this talk, I focus on what we have learned about collisionless shocks from the observations of GRB afterglows, and on the challenges to our theoretical understanding of the relevant physics. A brief introduction to GRBs and their afterglows is given in \S~\ref{sec:afterglow}. The dynamics of shock expansion during the afterglow phase is described in \S~\ref{sec:AGdynamics}, and the model for emission of afterglow radiation is discussed in  \S~\ref{sec:AGrad}. The collisionless nature of the shock is discussed in \S~\ref{sec:colles}. 

The theoretical challenges are discussed in \S~\ref{sec:challenge}. Our current understanding of, and open questions related to, the generation of magnetic fields are discussed in \S~\ref{sec:field}, where it is pointed out that the main challenge is the survival of the field at distances much larger than the skin depth downstream of the shock. Our current understanding of, and open questions related to, particle acceleration are discussed in \S~\ref{sec:particle}. One of the open questions related to particle acceleration is whether GRBs may produce the observed ultra-high energy (UHE), $>10^{19}$~eV, cosmic-rays. The challenges posed to all models of particle acceleration by the observations of UHE cosmic-rays, and the arguments suggesting GRBs may be the sources of these particles, are discussed in \S~\ref{sec:UHECR}. 

In the past few years, a significant effort was invested in 3D numerical studies of collisionless shocks. These studies are briefly discussed in \S~\ref{sec:numeric}, with an emphasis on what we have learned from the simulations, and on whether such simulation are likely to provide answers to the main open questions (for a more detailed summary of numerical simulation results see \cite{KKW06}). Recent theoretical results derived based on the assumption of a self-similar shock structure are briefly discussed in \S~\ref{sec:slfsim}. A brief summary is given in \S~\ref{sec:summary}.

It should be pointed out that although the discussion in this lecture is motivated by GRB afterglow observations, it may be relevant also for non relativistic collisionless shocks, such as shocks in young supernova remnants (SNRs) and in the intergalactic medium. In the past few years, high resolution X-ray observations have provided indirect evidence for the presence of strong magnetic fields, $\gtrsim 100\mu$G, in the \emph{non-relativistic} ($v/c\sim10^{-2}$) shocks of young SNRs (e.g. \cite{Volk05} and references therein). These fields extend to distances $D>10^{17}{\rm cm} \sim 10^{10}$~skin depths downstream, and possibly even $\gtrsim 10^{16}$~cm upstream, of the shock. In resemblance to GRBs, such strong
magnetic fields cannot result from the shock compression of a typical interstellar medium (ISM) magnetic field, $B_{\rm ISM}\sim \mbox{few }\mu$G. In SNRs the discrepancy is somewhat less severe, and the possibility that these magnetic fields are related to the large scale ISM fields cannot be ruled out. If the ISM magnetic fields can be neglected for SNR shocks, then much of the discussion presented below applies to such shocks as well.

\section{Gamma-ray bursts and their afterglow}
\label{sec:afterglow}

Gamma-ray bursts (GRBs) are short, typically tens of seconds long, flashes of gamma-rays, carrying most \begin{wrapfigure}{l}{80mm}
\includegraphics[width=80mm]{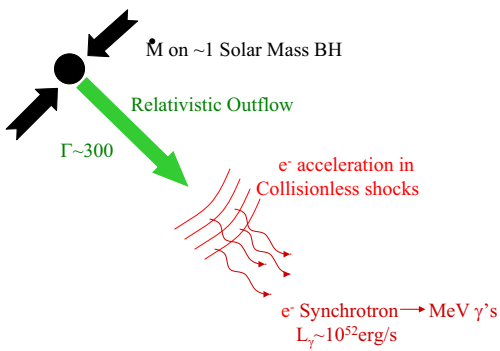}
\caption{The fireball scenario of GRB production}
\label{fig:GRB}
\end{wrapfigure}
of their energy in $>1$~MeV photons. The detection in the past few years of "afterglows," delayed X-ray, optical and radio emission from GRB sources, proved that the sources lie at cosmological distances, and provided strong support for the scenario of GRB production described in fig.~\ref{fig:GRB}. The energy source is believed to be rapid mass accretion onto a newly formed solar-mass black hole (or, possibly,  neutron star). Recent observations suggest that the formation of the central compact object is  associated with type Ib/c supernovae.

The energy release drives an ultra-relativistic plasma outflow, with Lorentz factor $\Gamma\sim10^{2.5}$. The emission of $\gamma$-rays is assumed to be due to internal collisionless shocks within the relativistic wind, the "fireball," which occur at a large distance from the central black-hole due to variability in the wind emitted from the central "engine." It is commonly assumed that electrons are accelerated to high energy within the collisionless shocks, and that synchrotron emission from these shock accelerated electrons produces the observed $\gamma$-rays. At still larger distances, the wind impacts on the surrounding medium. Here too, a collisionless shock driven into the ambient gas is believed to accelerate electrons to high energy, leading to synchrotron emission which accounts for the "afterglow." Afterglows are much better understood than the prompt $\gamma$-ray emission.

\subsection{Afterglow I: Dynamics}
\label{sec:AGdynamics}

As the fireball expands, it drives a relativistic shock (blast wave) into the surrounding gas, e.g. into the \begin{wrapfigure}{l}{70mm}
\includegraphics[width=70mm]{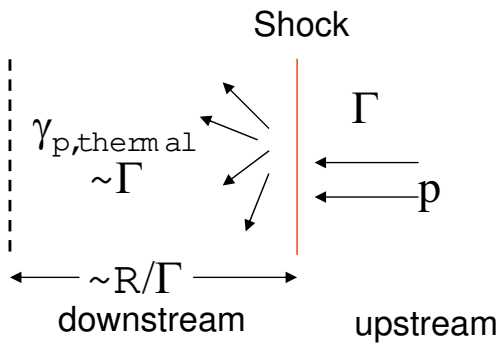}
\caption{The afterglow shock (shock frame).}
\label{fig:shock}
\end{wrapfigure}
interstellar medium (ISM) gas of the galaxy within which the explosion occurs. At early times, the fireball is little affected by the interaction with the ISM. When the mass of the shocked ISM plasma becomes significant, typically exceeding $E/\Gamma c^2$ where $E$ is the explosion energy, the fireball begins to decelerate. At sufficiently late time, when the shock radius becomes much larger than the radius of the onset of deceleration, most of the fireball energy is transferred to the ISM, and the flow approaches a self-similar behavior \cite{BnM76}. 

The dynamics of the shock can be easily understood by considering the flow in the shock frame, where the shock is stationary and the upstream plasma approaches it with high speed, corresponding to a Lorentz factor $\Gamma$. At the shock, protons are scattered and their momentum distribution is isotropized. The bulk velocity of the fluid is thus reduced, and a significant fraction of the incoming kinetic energy is converted to thermal energy. The characteristic Lorentz factor associated with the resulting thermal (i.e. random) motion of protons is  $\sim\Gamma$. In the observer frame, where the shock moves forward with Lorentz factor $\Gamma$, the energy of each proton is $\sim\Gamma^2m_pc^2$. The evolution of $\Gamma$ with shock radius $R$ is determined by energy conservation, $E\sim\Gamma^2(R) m_pc^2 n R^3$ where $n$ is the ambient medium number density. A more detailed analysis gives \citep{BnM76}
\begin{equation}
\Gamma(R) = \left(\frac{17E}{16\pi n m_p c^2}\right)^{1/2}R^{-3/2} = 
150\left({E_{53}\over n_0}\right)^{1/2}\left(\frac{R}{10^{17}{\rm cm}}\right)^{-3/2},
\label{eq:Gamma_BM}
\end{equation}
where $E = 10^{53}E_{53}$~erg and $n = 1n_0{\rm\ cm}^{-3}$ is the number density. $E_{53}\sim1$ for typical GRBs, and $n_0\simeq1$ for typical ISM. 

At the ISM rest frame, the shock reaches a radius $R$ at time $R/c$. The time measured in the shock frame is shorter by a factor $\Gamma$, $R/\Gamma c$. Since the plasma flows away from the shock (in the downstream region) at velocity $\sim c/3$ \citep{LL}, the shocked plasma is concentrated in a narrow shell of thickness $R/\Gamma$ (which becomes $R/\Gamma^2$ in the ISM rest frame due to Lorentz contraction), with proper density $\Gamma n$ (corresponding to a density $\Gamma^2n$ in the ISM frame). Finally, the radiation emitted by the shocked plasma at radius $R$ is delayed compared to that emitted $R=0$ by $R/c-R/v=R/2\Gamma^2c$, here $v\simeq(1-1/2\Gamma^2)c$ is the shock velocity (to leading order in $1/\Gamma^2$). A more detailed calculation gives \cite{WAG-ring} the following relation between the radius and the time measured by a distant observer: $t_{\rm obs.}(R) = R/4\Gamma(R)^2 c$, which implies
\begin{equation}
  R(t_{\rm obs.})=3.2\times10^{17}\left(\frac{E_{53}}{n_0}\right)^{1/4}\left(\frac{t}{1\rm week}\right)^{1/4}\,{\rm cm}.
\label{eq:R_obs}
\end{equation}

Eqs.~(\ref{eq:Gamma_BM}) and~(\ref{eq:R_obs}) provide a complete description of the dynamics of the afterglow shock. The self-similar model for shock expansion during afterglow emission has been tested by direct measurements of the fireball size, through diffractive scintillation \citep{WKF98} and through very large baseline interferometry \citep{Taylor04}, and by indirect size measurements using late time radio observations \citep{FWK00}. The simple self-similar model predictions are in excellent agreement with observations.

\subsection{Afterglow II: Radiation}
\label{sec:AGrad}

The agreement between the dynamics predicted by the blast wave model and the direct measurements of the fireball size strongly argue for the validity of this model's dynamics. As explained in \S~\ref{sec:colles}, the shock wave is most likely collisionless, i.e. mediated by plasma instabilities. The electromagnetic instabilities mediating the afterglow shock are expected to generate magnetic fields, and the rarity of binary particle collisions is expected to lead to the generation of a non-thermal particle distribution downstream \citep[see, e.g.][for a discussion of particle acceleration]{Blandford87}. Afterglow radiation was therefore predicted to result from synchrotron emission of shock accelerated electrons \cite{MnR97}. The observed spectrum of afterglow radiation is indeed remarkably consistent with synchrotron emission of electrons accelerated to a power-law distribution (see below), providing strong support to the validity of the standard afterglow model based on synchrotron emission of shock accelerated electrons \cite{WrevSN,Zhang04,Piran05}.

In order to determine the luminosity and spectrum of synchrotron radiation, the strength of the magnetic field and the energy distribution of the electrons must be determined. Due to the lack of a first principles theory of collisionless shocks, we present in this section a purely phenomenological approach to the model of afterglow radiation emission. That is, we do not discuss the processes responsible for particle acceleration and magnetic field generation. Rather, we simply assume that a fraction $\epsilon_B$ of the post-shock thermal energy density is carried by the magnetic field, that a fraction $\epsilon_e$ is carried by electrons, and that the energy distribution of the electrons is a power-law, $d\log n_e/d\log\varepsilon=p$ (above some minimum energy $\varepsilon_0$ which is determined by $\epsilon_e$ and $p$). $\epsilon_B,\,\epsilon_e$ and $p$ are treated as free parameters, to be determined by observations. It is important to clarify here that the constraints implied 
on these parameters by observations are independent of any assumptions regarding the nature of the afterglow shock and the processes responsible for particle acceleration or magnetic field generation. Any model proposed for the shock should satisfy these observational constraints.

The parameters $\epsilon_B,\,\epsilon_e$ and $p$, together with the parameters $E$ and $n$ which determine the shock dynamics, completely determine the magnetic field strength and electron distribution (including their temporal and spatial dependence). Thus, they completely determine the luminosity and spectrum of synchrotron emission, including its time dependence. The power-law dependence of $\Gamma$ on time, combined with the power-law dependence of the number of electrons on energy, imply that the time dependence of the synchrotron emission (at a given frequency) and its frequency dependence (at a given time) are both power-laws. The indices of these power-laws are both determined by $p$ alone. One of the most impressive successes of the model described here for afterglow emission is that the observed frequency and time dependence of the afterglow flux both follow power-laws, and that the temporal and frequency power-law indices are both consistent with the same value of $p$. Moreover, the inferred value of $p$, $p\approx2$, is consistent with the theoretical expectations for particle acceleration in collisionless shocks (this is discussed in detail in \S~\ref{sec:particle}).

In addition to $p$, the afterglow model is determined by 4 parameters, $\{E,n,\epsilon_B,\epsilon_e\}$. In principle, all 4 parameters may be determined by observations, which provide 4 observables. The synchrotron spectrum can be described as a combination of 4 power-laws, with 3 break frequencies: the frequency $\nu_a$ below which the synchrotron self-absorption optical depth exceeds unity, the frequency $\nu_m$ where the flux peaks (corresponding to the synchrotron frequency of electrons with energy $\varepsilon_0$), and the frequency $\nu_c$ corresponding to synchrotron frequency of electrons at energy $\varepsilon_c$ for which the synchrotron cooling time is comparable to the dynamical time (the time for significant shock expansion). Typically, $\nu_a<\nu_m<\nu_c$, and the specific intensity $f_\nu$ is a broken power-law, $f_\nu\propto\nu^\alpha$ with $\alpha={2,1/3,-(p-1)/2,-p/2}$ from low to high frequency. Afterglow observations therefore provide 4 observables, $\nu_a\,,\nu_m\,,\nu_c$ and the normalization of the flux, e.g. $f_\nu(\nu=\nu_m$). This, in turn, allows one to determine the 4 model parameters, $\{E,n,\epsilon_B,\epsilon_e\}$.

As noted above, the values which are typically obtained for $E$ and $n$ are $E\sim10^{53}$~erg and $n\sim1{\rm cm}^{-3}$. The inferred energy is consistent with that expected based on $\gamma$-ray observations, and the inferred density is consistent with that typically expected in the ISM. Here we are more interested in the inferred values of $\epsilon_B,\,\epsilon_e$. It is natural to hope that the values of $\epsilon_B$, $\epsilon_e$ (and $p$) are universal since they are determined by the microphysics of the collisionless shock. The constancy of $p$ and of $\epsilon_e$ among different bursts is strongly supported by observations. Universal values of $p$ and $\epsilon_e$, $p\approx2$ and $\epsilon_e\approx0.1$, typically inferred from most optical afterglows, are also inferred from the clustering of explosion energies \citep{Frail01} and from X-ray afterglow
luminosity \citep{Freedman01,Berger03}. The value of $\epsilon_B$ is less well constrained by observations. However, in cases where $\epsilon_B$ can be reliably constrained by multi waveband spectra, values close to equipartition, $\epsilon_B=0.01$ to $0.1$, are inferred \citep[e.g.][]{FWK00}. Observations are consistent with the values of $\epsilon_B$, $\epsilon_e$ and $p$ being independent of $\Gamma$.

It should be pointed out here that \citet{Eichler05} have shown that observations determine $\epsilon_e$ and $\epsilon_B$ (and also $E$ and $n$) only up to a factor $f$, the fraction of electrons accelerated, where $m_e/m_p<f<1$. However, it is expected that $f$ is not very small, $f\gtrsim 1/10$ \citep{Eichler05}.

\subsection{Why collisionless shock?}
\label{sec:colles}

As explained in the preceding sub-sections, afterglow observations strongly support a model where the afterglow is produced by a relativistic shock wave driven by the explosion into the surrounding medium, and where the radiation is due to synchrotron emission of high energy electrons, following a power-law energy distribution, radiating in the post-shock magnetic field. These conclusions are independent of assumptions related to the nature of the shock and to the processes responsible for particle acceleration and magnetic field generation. The nature of the shock is the main subject discussed in this sub-section.

The scattering of particles at the shock, leading to isotropization of momenta in the downstream, can not be mediated by binary particle collisions. Let us first consider shock mediation through binary Coulomb collisions. At the shock frame, the number density of particles is $\Gamma n$ (both in the up- and downstream), and the cross section for Coulomb collisions may be estimated as $\pi d^2$ with $d\sim e^2/\Gamma m_p c^2$. The resulting mean-free-path, $\lambda\sim 1/\Gamma n\pi d^2\sim10^{31}\Gamma n_0^{-1}$~cm, is many orders of magnitude larger than the size of the system, $\sim10^{17}$~cm (in fact, it is larger than the size of the observable universe, $\sim10^{28}$~cm). Coulomb collisions can not therefore mediate the shock. The mean free path for nuclear collisions, $\lambda\sim 1/\Gamma n\sigma_{pp}\sim10^{25}(\Gamma n_0)^{-1}$~cm, is also many orders of magnitude larger than the size of the system. This implies that nuclear collisions can not mediate the shock either. Note, that we are considering here the scattering of protons (rather than electrons), since they carry most of the momentum of the incoming plasma.

The scattering of particles at the shock is most likely mediated by collective plasma instabilities (involving a macroscopic number of particles). As the fast, expanding downstream plasma tries to propagate through the upstream plasma, which is at rest, electromagnetic instabilities develop, generating electric and magnetic fields in the plasma, which deflect the particles and tend to isotropize their momentum distribution. These instabilities develop on a time scale comparable to the inverse of the (relativistic) plasma frequency, $t\sim1/\omega_p$. Since the particle distribution is expected to be isotropized on a time scale of $1/\omega_p$, the width of the shock (the size of the region over which particles are being scattered) is expected to be of the order of several skin depths, few times $c/\omega_p$, where
\begin{equation}\label{eq:skindepth}
    c/\omega_p=c(4\pi\Gamma n e^2/\Gamma m_p)^{-1/2}=c(4\pi n e^2/ m_p)^{-1/2}\sim10^7 n_0^{-1/2}\,{\rm cm}.
\end{equation}
Note, that in the shock frame both the number density and the "effective" mass of protons are larger by a factor $\Gamma$ than their values in the ISM, which implies that the plasma frequency is similar to that of the ISM. 

A shock which is mediated by collective plasma instabilities, instead of by binary particle collisions, is termed "collisionless." Afterglow shocks, as shocks in many other astrophysical systems, are collisionless due to the low densities characteristic of these systems.

The following comment is in place at this point. The arguments presented above indicate that a collisionless shock would be formed, with a width of the order of several skin depths. This conclusion is supported by numerical simulations (see \S~\ref{sec:numeric} for discussion). It should however be kept in mind that a self-consistent theory of collisionless shocks does not yet exist, and that the nature of the shock may thus be different than described above. In this context it is important to clarify that the conclusions inferred in \S~\ref{sec:AGdynamics} regarding the existence of a blast wave and its dynamics, and in \S~\ref{sec:AGrad} regarding the strength of the post shock magnetic field and the electron energy distribution, are inferred directly from afterglow observations and are independent of any assumptions regarding the nature of the shock and the processes responsible for particle acceleration or magnetic field generation. One may not rule out, of course, the possibility of explaining afterglow observations with a different model, where a shock wave does not form or where radiation is not produced by synchrotron emission. Such an alternative models have not, however, been put forward yet.

\section{Challenges}
\label{sec:challenge}

\subsection{Magnetic field generation}
\label{sec:field}

Afterglow shocks are highly "non-magnetized:" The ratio of magnetic field to kinetic energy flux ahead of the shock is very small, $U_{B,up}/nm_pc^2\sim 10^{-10}$, where $U_{B,up}$ is the magnetic energy density in the upstream, and $B_{up}\sim3\mu$~G is typical to the ISM. This strongly suggests that the shock structure is determined by the upstream density and the shock Lorentz factor alone \citep[e.g.][]{Gruzinov01a}. It is reasonable to assumed that the upstream magnetic field does not play a role in the determination of the shock structure, since the ratio of the (relativistic) cyclotron frequency of the thermal protons, $\omega_L\sim eB/\Gamma m_pc$, to the plasma frequency is $(\omega_L/\omega_p)^2\sim U_{B,up}/nm_pc^2\sim 10^{-10}$, which implies that thermal protons in the shock frame are not affected by the compressed upstream magnetic field on a the dynamical time $\sim\omega_p^{-1}$.

The downstream magnetic field, implied by afterglow observations, is close to equipartition, i.e. it's energy density is higher than that of the upstream field by a factor of $\sim10^8$. Such near equipartition magnetic field may conceivably be produced in the collisionless shock driven by the GRB explosion by electromagnetic (e.g. Weibel) instabilities \citep[e.g.][]{Blandford87,Gruzinov99,Medvedev99,Wiersma04}.
The main challenge associated with the downstream magnetic field is related to the fact that in order to account for the observed radiation as synchrotron emission from accelerated electrons, the field amplitude must remain close to equipartition deep into the downstream, over distances $\sim10^{10}c/\omega_p$: At $t\sim1$~day the magnetic field must be strong throughout the (proper) width $\sim2\Gamma c t\sim R/\Gamma\sim10^{17}$~cm while $c/\omega_p\sim10^7n_0^{-1/2}$~cm. This is a challenge since electromagnetic instabilities are believed to generate (near-equipartition)
magnetic fields with coherence length $L\sim c/\omega_p$, and a field varying on such scale is expected to decay within a few skin-depths downstream \citep{Gruzinov01a}. This suggests that the correlation length of the magnetic field far downstream must be much larger than the skin depth, $L\gg c/\omega_p$, perhaps even of the order of the distance from the shock \citep{Gruzinov99,Gruzinov01a}.

\subsection{Particle acceleration}
\label{sec:particle}

The mechanism which is commonly believed to be responsible for the production of non-thermal distributions of high energy particles in many astronomical systems (e.g. planetary bow shocks within the solar wind, supernovae remnant shocks, jets of radio galaxies, GRB's and possibly shocks involved in the formation of the large scale structure of the universe) is the diffusive (Fermi) acceleration of charged particles in collisionless shocks. In this process, high energy particles are scattered back and forth across the shock discontinuity. Since the flow at the shock is converging (that is, the fluid on each side of the shock discontinuity "sees" the fluid at the other side as approaching it), each time the particle is scattered from one side of the shock to the other it gains energy. Since the scattering is mediated by interactions with plasma waves, incorporating a macroscopic number of particles, the scattered particle can reach very high energies. One may consider an analogy where a light ping-pong ball is scattered back and forth between approaching trains, gradually gaining energy with each collision. The accelerated particle gains energy until it escapes the shock at the far downstream.

Despite decades of research, this mechanism is still not understood from first principles (see, e.g., \cite{arons,Dieckmann06a} for discussions of alternative shock acceleration processes). Particle scattering in collisionless shocks is due to electro-magnetic waves. No present analysis self-consistently calculates the generation of these waves, the scattering of particles and their acceleration. Most analyses consider, instead, the evolution of the particle distribution adopting some Ansatz for the particle scattering mechanism (e.g. diffusion in pitch angle), and the "test particle" approximation, where modifications of shock properties due to the high energy particle distribution are neglected \citep[see, however][]{Ellison02}. 

This phenomenological approach proved successful in accounting for non-thermal particle distributions inferred from observations. The theory of diffusive particle acceleration in \emph{non-relativistic} shocks was first developed in \cite{Krymskii77, Axford77, Bell78, Blandford78}, and was shown to predict a power-law distribution of particle momenta, $dn/d\varepsilon\propto\varepsilon^{-p}$, with 
\begin{equation}\label{eq:p_NR}
    p=\frac{3\beta_u}{\beta_u-\beta_d}-2.
\end{equation}
\begin{wrapfigure}{l}{70mm}
\includegraphics[width=70mm]{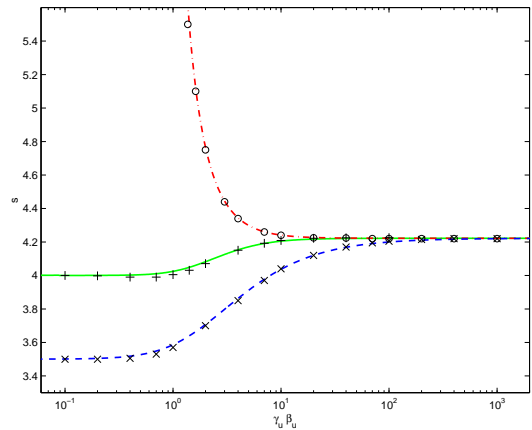}
\caption{Accelerated particle spectral index, $s=p+2$, as a function of $\Gamma v/c$ (where $\Gamma$ and $v$ are the shock Lorentz factor and speed) for various equations of state of the fluid. Solid lines show the analytic result, eq.~(\ref{eq:p_R}), and symbols show the results of numerical simulations.}
\label{fig:keshet}
\end{wrapfigure}
Here $\beta_u$ ($\beta_d$) is the upstream (downstream) fluid velocity normalized to the speed of light. For strong shocks in an ideal gas of adiabatic index $\hat{\gamma}=5/3$, this implies $p=2$, in agreement with observations of non-relativistic shocks.

Observations of GRB afterglows lead to the conclusion that the highly relativistic collisionless afterglow shocks produce a power-law distribution of high energy particles with $p=2.2\pm0.2$ \cite{grb_s,Freedman01,Berger03}. This triggered a numerical investigation of particle acceleration in such shocks \cite{Bednarz98}. The index $p$ was calculated under the "test particle" approximation for a wide range of Lorentz factors and various equations of state \cite[][and references therein] {Bednarz98, Kirk00, Achterberg01}. In particular, $p$ was shown to approach the value $2.2$ for large Lorentz factors, in accord with GRB observations. These studies have assumed rest frame diffusion in pitch angle \cite{Bednarz98} or in the angle between particle velocity and shock normal \cite{Kirk00, Achterberg01}. These two assumptions may not be valid, for example if large angle scattering prevails \cite{Meli03}. Both yield similar spectra for ultra-relativistic shocks \cite{Ostrowski02}. 

Recently, an analytical study of diffusive particle acceleration in relativistic, collisionless shocks yielded a simple relation between the spectral index $p$ and the anisotropy of the momentum distribution along the shock front \citep{Keshet05}. Based on this relation, a generalization of eq.~(\ref{eq:p_NR}) was derived for relativistic shocks, under the assumption of isotropic diffusion, 
\begin{equation}\label{eq:p_R}
    p=\frac{3\beta_u-2\beta_u\beta_d^2+\beta_d^3}{\beta_u-\beta_d}-2.
\end{equation}
This result is in agreement with previous numerical determinations of $p$ for all shock speeds and fluid equations of state (see fig.~\ref{fig:keshet}). In particular, it yields $p=20/9=2.22$ in the ultra-relativistic limit. 

The results described above indicate that we have a good understanding of particle acceleration in relativistic shocks in the test particle approximation. Several major open questions remain. First, the spectral index $p$ depends on the form of the diffusion function (unlike the situation in non relativistic shocks), and a value consistent with observations, $p\approx2$, is obtained for isotropic diffusion only. Second, the fraction of particles which are accelerated can not be determined. Finally, and most importantly, a theory that self-consistently describes the acceleration of particles and the generation of electromagnetic waves is still missing. The accelerated particles are estimated to carry a considerable part of the energy: electrons alone carry $\sim 10\%$ of the energy in GRB afterglow shocks and $\sim 5\%$ of the energy in SNR shocks \citep[][and references therein]{Keshet04}, and at least $10\%$ of the energy in SNR shocks must be converted into relativistic protons if these shocks are responsible for Galactic cosmic rays \citep[e.g.][]{Drury89}. This implies that the accelerated particles are likely to have an important role in generating and maintaining the inferred magnetic fields. This conclusion is supported also by the evidence of strong amplification of the magnetic field in the upstream of GRB afterglow shocks \cite{Zhuo06}, which is most likely due to the streaming of high energy particles ahead of the shock. Since the high energy particles are likely to play an important role in the generation of the fields, a theory of collisionless shocks must provide a self-consistent description of particle acceleration, which depends on the scattering of these particles by magnetic fields, and field generation, which is likely driven by the accelerated particles.

\subsection{High energy cosmic rays}
\label{sec:UHECR}

The cosmic-ray spectrum extends to energies $\sim10^{20}$~eV \cite{CR_data_rev}, and is likely dominated beyond $\sim10^{19}$~eV by extra-Galactic sources of protons. The origin of the ultra high energy (UHE), $>10^{19}$~eV, cosmic rays is a mystery: The stringent constraints, which are imposed on the properties of possible UHE cosmic ray sources by the high energies observed, rule out almost all source candidates. The essence of the challenge of accelerating particles to $>10^{19}$~eV can be understood using the following simple arguments, which are independent of the details of the acceleration model \cite{WNobel04}. 

\begin{wrapfigure}{l}{70mm}
\includegraphics[width=70mm]{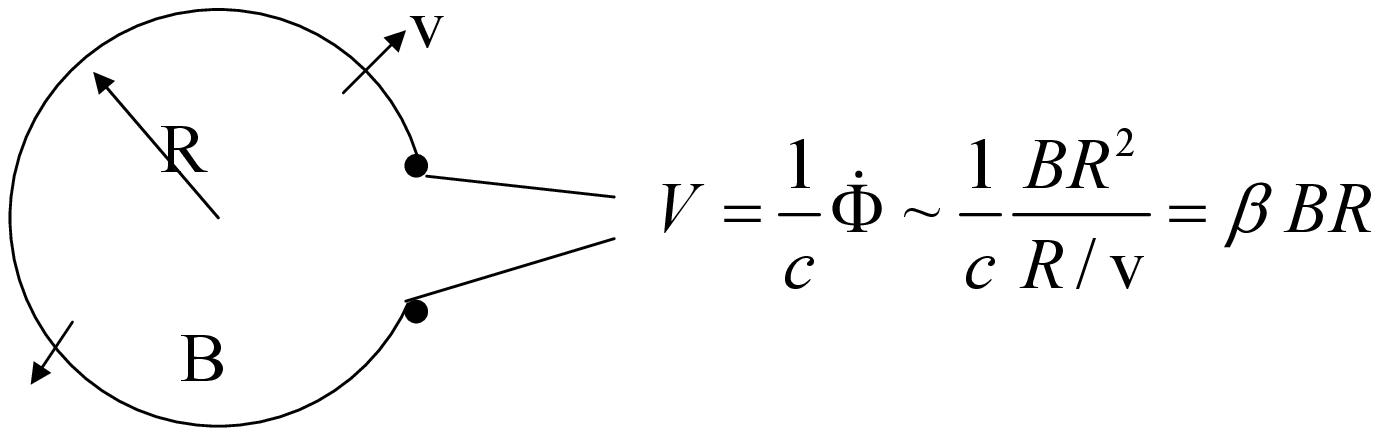}
\caption{Potential drop generated by a non-steady outflow of magnetized plasma.}
\label{fig:acceleration}
\end{wrapfigure}

Consider an astrophysical source driving a flow of magnetized plasma, with characteristic magnetic field strength $B$ and velocity $v$. Imagine now a conducting wire encircling the source at radius $R$, as illustrated in fig.~\ref{fig:acceleration}. The potential generated by the moving plasma is given by the time derivative of the magnetic flux $\Phi$ and is therefore given by $V=\beta B R$ where $\beta=v/c$. A proton which is allowed to be accelerated by this potential drop would reach energy
$\varepsilon_p\sim\beta eB R$. The situation is somewhat more complicate in the case of a relativistic outflow, where $\Gamma\equiv(1-\beta^2)^{-1/2}\gg1$. In this case, the proton is allowed to be accelerated only over a fraction of the radius $R$, comparable to $R/\Gamma$. To see this, one must realize that as the plasma expands, its magnetic field decreases, so the time available for acceleration corresponds, say, to the time of expansion from $R$ to $2R$. In the observer frame this time is $R/c$, while in the plasma rest frame it is $R/\Gamma c$. Thus, a proton moving with the magnetized plasma can be accelerated over a transverse distance $\sim R/\Gamma$. This sets a lower limit to the product of the magnetic field and source size, which is required to allow acceleration to $\varepsilon_p$,
\begin{equation}\label{eq:BR}
    BR>\Gamma \varepsilon_p/e\beta.
\end{equation}

Eq.~\ref{eq:BR} also sets a lower limit to the rate $L$ at which energy should be generated by the source. The magnetic field carries with it an energy density $B^2/8\pi$, and the flow therefore carries with it an energy flux $>vB^2/8\pi$ (some energy is carried also as plasma kinetic energy), which implies $L>vR^2B^2$. Using eq.~\ref{eq:BR} we find
\begin{equation}\label{eq:L}
  L>\frac{\Gamma^2}{\beta}\left(\frac{\varepsilon_p}{e}\right)^2c
  =10^{45.5}\frac{\Gamma^2}{\beta}\left(\frac{\varepsilon_p}{10^{20}\rm eV}\right)^2{\rm erg/s}.
\end{equation}
Only two types of sources are known to satisfy this requirement. The brightest steady sources are active galactic nuclei (AGN). For them $\Gamma$ is typically between 3 and 10, implying $L>10^{47}{\rm erg/s}$, which may be satisfied by the brightest AGN. The brightest transient sources are GRBs. For these sources $\Gamma\simeq10^{2.5}$ implying $L>10^{50.5}{\rm erg/s}$, which is generally satisfied since the typical observed MeV-photon luminosity of these sources is $L_\gamma\sim10^{52}{\rm erg/s}$. For a more detailed discussion of the arguments suggesting an association of GRBs and UHE cosmic-rays see \citep{WPylos06} and references therein.

\section{Numerical simulations}
\label{sec:numeric}

The search for a self-consistent theory of collisionless shocks has led to extensive numerical studies. Particle in cell (PIC) simulations were performed in one dimension \citep[e.g.][]{Dieckmann06}, in two dimensions \citep[e.g.][]{Wallace91,Kato05}, in two spatial and three velocity dimensions \citep[2D3V; e.g.][]{Gruzinov01a,Gruzinov01b,Medvedev05} and in the last few years also in three dimensions \citep{Nishikawa03,Frederiksen04,Spitkovsky05}. In addition to shock simulations, homogeneous 3-dimensional simulations of inter-penetrating plasmas have been performed \citep{Silva03,Jaroschek04,Romanov04}. It is important to note here that the applicability of 1-dimensional and 2-dimensional simulations to the problem of collisionless shocks is questionable, since the 1- or 2-dimensional symmetry imposed prevents the evolution of 3-dimensional modes, which are likely important for the generation of the shock structure. Similarly, the applicability of homogeneous (inter-penetrating) plasma simulations is questionable, since the process of particle acceleration, which is likely related to scattering of particles across the shock and which may play an important role in generating large scale magnetic fields, may be completely absent in such simulation.

The 3-dimensional simulations have provided compelling evidence that transverse electromagnetic (Weibel- like) instabilities can mediate relativistic collisionless shocks, and that the upstream magnetic field is unimportant if the upstream is not highly magnetized. As expected, the width of the shock transition region is found to be a few $\times10~c/\omega_p$ \citep[see, e.g.,][]{Spitkovsky05}. These simulations also indicate that near-equipartition magnetic fields are generated in the downstream of relativistic shocks propagating in pair ($e^+e^-$) plasmas, and $\epsilon_B\ge m_e/m_p$ magnetic fields are generated in the downstream of shocks in electron-proton plasmas. Simulations of electron-proton plasmas are forced to employ an effective small proton to electron mass ratio, $\tilde{m}_p/m_e\lesssim 20$ with present computational resources, and the preliminary results thus obtained are not easily extrapolated to more realistic mass ratios. 

3D simulations are limited to very small simulation boxes, and can reliably probe small length scales no larger than $\sim 100$ electron skin-depths, and short time scales no longer than $\sim 100$ electron plasma times. Obviously, the question of field survival and correlation length evolution on length scales $\gg c/\omega_p$, which is the main challenge to our theoretical understanding, are not yet answered.  Similarly, highly energetic particles cannot be contained in the small simulation boxes used, so Fermi-like acceleration processes are suppressed. It appears, therefore, that the numerical resources required to answer some of the main open questions by direct 3-dimensional numerical simulations will not be available in the near future. 

It is also important to note here, that some published results are based on PIC simulations in stages where the boundary conditions strongly modify the plasma evolution. For example, claims that the magnetic fields decay slowly or saturate at some finite level remain questionable, until verified by simulations with sufficiently large simulation boxes.

\section{Self-similarity}
\label{sec:slfsim}

As discussed in \S~\ref{sec:field}, afterglow observations suggest that the characteristic length scale $L$ for variations in the magnetic field becomes much larger than the skin depth, $L\gg c/\omega_p$, at distances $D\gg c/\omega_p$ downstream of the shock. It is thus reasonable to assume that, at $D\gg c/\omega_p$, $L$ is the only relevant length scale, which implies self-similarity \citep{KKW06}. There is no proof that self-similarity will be present whenever the characteristic length scale diverges (or becomes infinitesimal). However, the self-similarity assumption is known to be valid for many physical systems in which such divergence occurs (see, e.g., \cite{zel68} for self-similarity in hydrodynamics, and \cite{Kadanoff67} for self-similarity in critical phenomena). 

Self-similarity implies that the plasma configurations at different distances $D$ downstream of the shock, corresponding to different values of $L$, are similar to each other. Consider for example the average of the particle distribution function over planes perpendicular to the shock normal, $<f_\alpha>(D,\mathbf{p})$, or the magnetic field correlation function, $B_{ij}(\Delta\mathbf{x},\Delta t,D)\equiv<B_i(\mathbf{x},t)B_j(\mathbf{x}+\Delta\mathbf{x},t+\Delta t)>$ (Here, $\mathbf{p}$ is the particle momentum, $<>$ denotes an average over planes perpendicular to the shock normal, and we have assumed that such averages depend only on $D$). Self-similarity implies that such averages scale with $L$, i.e. that 
\begin{align}
<{f_{\alpha}}>(\mathbf{p},L)=
\left(\frac{L}{L_0}\right)^{s_f}
<{f_\alpha}>\left[\frac{\mathbf{p}}{(L/L_0)^{s_p}},L_0\right],
\end{align}
\begin{align}
 B_{ij}(\Delta\mathbf{x},\Delta t,L)=\left(\frac{L}{L_0}\right)^{2s_B}
B_{ij}\left[\frac{\Delta\mathbf{x}}{L/L_0},\frac{\Delta t}{(L/L_0)^{s_t}},L_0\right].
\end{align}
Here, we have replaced the dependence on $D$ with a dependence on $L$, as the self-similarity assumption implies
\begin{equation}\label{eq:LD}
    L\propto D.
\end{equation}

Using the Maxwell-Vlasov equations, the similarity indices may be derived \citep{KKW06}:
\begin{equation}
-1< s_B\le0,\quad s_t=1,\quad s_p=s_B+1,\quad s_f=-4-2s_B.
\end{equation}
These relations imply that the characteristic Larmor radius of energetic particles scales as $L$, and that the energy density of energetic particles in any momentum interval, with the interval scaling as $L^{s_p}$, scales as the magnetic field energy density $\propto L^{2s_B}$. Under the assumption that accelerated particles reach the shock front (and/or are advected to the downstream), the spectrum of accelerated particles must follow
\begin{equation}
dn/d\varepsilon\propto \varepsilon^{-2/(s_B+1)}.
\end{equation}

The self-similarity assumption suggests that the plasma may be approximately described as a combination of two self-similar components: a kinetic component of energetic particles, and an MHD-like component representing "thermal" particles. It is likely that the thermal component may be considered as an infinitely conducting fluid, in which case $s_B=0$ and the scalings are completely determined, e.g. $dn/d\varepsilon\propto \varepsilon^{-2}$ and $B\propto D^0$, with possible logarithmic corrections. 

The self-similarity assumptions may be tested through their predictions for the evolution of homogenous (time-dependent) plasmas, which may be accessible to direct numerical simulations. An inclusion (at the initial conditions) of a non isotropic, power-law spectrum of high energy particles, $dn/d\varepsilon\propto \varepsilon^{2+l_p}$, in homogenous simulations may lead to a self-similar evolution in time, with magnetic field evolution following $B\propto t^{-(l_p+4)/2(l_p+3)}$.

\section{Summary}
\label{sec:summary}

\begin{enumerate}
    \item GRB afterglows provide a unique probe of relativistic collisionless shocks, with Lorentz factors in the range of $\sim100$ to $\sim1$.
    \item Afterglow shock ares "non-magnetized," in the sense that the ratio of magnetic field to kinetic energy flux ahead of the shock is very small, $U_{B,up}/nm_pc^2\sim 10^{-10}$, where $U_{B,up}$ is the magnetic energy density in the upstream. This suggests that the upstream field plays no role in determining the shock structure.
    \item The downstream magnetic field implied by afterglow observations is close to equipartition, i.e. it's energy density is higher than that of the upstream field by a factor of $\sim10^8$. Such near equipartition magnetic field may conceivably be produced by electromagnetic (Weibel- like) instabilities.
    \item The main challenge associated with the downstream magnetic field is related to the fact that in order to account for the observed radiation as synchrotron emission from accelerated electrons, the field amplitude must remain close to equipartition deep into the downstream, over distances $\sim10^{10}c/\omega_p$. This is a challenge since electromagnetic instabilities are believed to generate (near-equipartition) magnetic fields with coherence length $L\sim c/\omega_p$, and a field varying on such scale is expected to decay within a few skin-depths downstream. This suggests that the correlation length of the magnetic field far downstream must be much larger than the skin depth, $L\gg c/\omega_p$, perhaps even of the order of the distance from the shock. 
    \item Observations of GRB afterglows imply that relativistic collisionless shocks produce a power-law distribution of high energy particles, $d\log n/d\log\varepsilon=-2.2\pm0.2$. This is consistent with the results of diffusive (Fermi) acceleration of test particles, under the assumption of isotropic diffusion. However, a theory that self-consistently describes the acceleration of particles and the generation of electromagnetic waves is still missing. Since the high energy particles are likely to play an important role in the generation of the fields, a theory of collisionless shocks must provide a self-consistent description of particle acceleration, which depends on the scattering of these particles by magnetic fields, and field generation, which is likely driven by the accelerated particles.
    \item One of the important questions related to particle acceleration in GRBs is whether thes objects may produce the observed UHE, $>10^{19}$~eV cosmic-rays. While there is evidence for an association between GRB and UHE cosmic-ray sources, the model for particle acceleration is incomplete.
    \item The search for a self-consistent theory of collisionless shocks has led to extensive numerical studies. 3D simulations provide compelling evidence that transverse electromagnetic instabilities can mediate relativistic collisionless shocks, that the upstream magnetic field is unimportant if the upstream is not highly magnetized, that the width of the shock transition region is a few $\times10~c/\omega_p$, and that near-equipartition magnetic fields are generated in the downstream of relativistic shocks. 
    \item These simulations are limited, however, to very small simulation boxes, and can reliably probe small length scales no larger than $\sim 100$ electron skin-depths, and short time scales no longer than $\sim 100$ electron plasma times.  Obviously, the question of field survival and correlation length evolution on length scales $\gg c/\omega_p$, which is the main challenge to our theoretical understanding, are not yet answered. Similarly, highly energetic particles cannot be contained in the small simulation boxes used, so Fermi-like acceleration processes are suppressed. 
    \item The growth of the magnetic field correlation length to scales much larger than the skin depth suggests that downstream shock structure is self-similar. Assuming self-similarity allows one to draw important conclusion regarding the shock structure. In particular, the the magnetic field amplitude should scale with the distance $D$ from the shock as $B\propto D^{s_B}$ with $-1<s_B\le0$, and the particle energy distribution should be a power-law, $dn/d\varepsilon\propto \varepsilon^{-p}$ with $p=2/(s_B+1)$ (up to logarithmic corrections). If the low energy particles can be considered as infinitely conducting, the scaling is completely determined,  $s_B=0$ and $p=2$. The self-similarity assumption may be tested by homogeneous plasma simulations.

\end{enumerate}    

\acknowledgements
Research supported in part by an ISF grant.


\begin{thebibliography}{99}

\bibitem[{{Blandford} \& {Eichler}(1987)}]{Blandford87}
{Blandford}, R. \& {Eichler}, D. 1987, Phys. Rep. 154, 1

\bibitem[{{Begelman} {et~al.}(1994)}]{Begelman94}
{Begelman}, M.~C., {Rees}, M.~J.,  \& {Sikora}, M. 1994, \apj,
429, L57

\bibitem[{{Maraschi}(2003)}]{Maraschi03}
{Maraschi}, L. 2003, in AGNs: from Central Engine to Host Galaxy,
Eds. S. Collin, F. Combes \& I. Shlosman. ASP, Conference Series,
{ 290}, 275

\bibitem{WrevSN} 
  Waxman, E. 2003, in \emph{Supernovae and Gamma-Ray bursters}, Ed. K. Weiler (Springer), Lecture Notes in 
  Physics 598, 393 (arXiv:astro-ph/0303517)

\bibitem[{{Zhang} \& {M{\'e}sz{\'a}ros}(2004)}]{Zhang04}
{Zhang}, B. \& {M{\'e}sz{\'a}ros}, P. 2004, International Journal of
Modern Physics A, 19, 2385

\bibitem[Piran (2005)]{Piran05} 
 Piran, T.\ 2005, Rev. Mod. Phy. 76, 1143

\bibitem[{{Loeb} \& {Waxman}(2000)}]{Loeb00}
{Loeb}, A. \& {Waxman}, E. 2000, Nature, 405, 156

\bibitem[Gruzinov (2001)]{Gruzino01}
 Gruzinov, A. 2001, \apj, 563, L15

\bibitem[Schlickeiser \& Shukla (2003)]{Schlick03}
  Schlickeiser, R. \& Shukla, P. K. 2003, \apj, 599, L57

\bibitem[{{Krall}(1997)}]{Krall97}
{Krall}, N.~A. 1997, Advances in Space Research, 20, 715

\bibitem[Katz, Keshet \& Waxman (2006)]{KKW06}
  Katz, B., Keshet, U. \& Waxman, E. 2006, arXive:astro-ph/0607345

\bibitem[{{V{\"o}lk} {et~al.}(2005){V{\"o}lk}, {Berezhko}, \&
  {Ksenofontov}}]{Volk05}
{V{\"o}lk}, H.~J., {Berezhko}, E.~G., \& {Ksenofontov}, L.~T. 2005,
A\&A, 433,
  229

\bibitem[Blandford \& Mckee (1976)]{BnM76} 
  Blandford, R. D. \& Mckee, C. F. 1976, Phys. Fluids 19, 1130

\bibitem[Landau \& Lifshitz (1987)]{LL}
  Landau, L. D. \& Lifshitz, E. M. 1987, {\it Fluid mechanics} (Pergamon Press, 1987). 

\bibitem[Waxman (1997)]{WAG-ring} 
  Waxman, E. 1997, Ap. J. 491, L19

\bibitem[Waxman, Kulkarni \& Frail~(1998)]{WKF98} 
  Waxman, E., Kulkarni, S. R. \& Frail, D. A. 1998, \apj {497}, 288 
  
\bibitem[Taylor et al.~(2004)]{Taylor04}
  Taylor, G. B., Frail, D. A., Berger, E. \& Kulkarni, S. R. 2004, \apj 609, L1

\bibitem[Frail, Waxman \& Kulkarni~(2000)]{FWK00} 
  Frail, D. A., Waxman, E. \& Kulkarni, S. R. 2000, \apj 537, 191
  
\bibitem[M\'esz\'aros \& Rees~(1997)]{MnR97} 
  M\'esz\'aros, P. \& Rees, M. J. 1997, \apj 476, 232

\bibitem[Frail et al. (2001)]{Frail01} 
  Frail, D. A.  et al. 2001, Ap. J. 562, L55

\bibitem[Freedman \& Waxman (2001)]{Freedman01} 
  Freedman, D. L. \& Waxman, E. 2001, Ap. J. 547, 922

\bibitem[Berger, Kulkarni \& Frail (2003)]{Berger03} 
  Berger, E., Kulkarni, S. R. \& Frail, D. A. 2003, Ap. J. 590, 379

\bibitem[{{Eichler} \& {Waxman}(2005)}]{Eichler05}
{Eichler}, D. \& {Waxman}, E. 2005, Ap. J., 627, 861

\bibitem[Gruzinov \& Waxman (1999)]{Gruzinov99} 
 Gruzinov, A. \& Waxman, E. 1999, Ap. J. 511, 852

\bibitem[{{Medvedev} \& {Loeb}(1999)}]{Medvedev99}
{Medvedev}, M.~V. \& {Loeb}, A. 1999, Ap. J. 526, 697

\bibitem[{{Wiersma} \& {Achterberg}(2004)}]{Wiersma04}
{Wiersma}, J. \& {Achterberg}, A. 2004, A\&A 428, 365

\bibitem[{{Gruzinov}(2001)}]{Gruzinov01a}
{Gruzinov}, A. 2001, Ap. J. 563, L15

\bibitem[Arons \& Tavani (1994)]{arons}
  Arons, J. \& Tavani, M. 2004, Ap. J. Supp. Series {\bf 90}, 797

\bibitem[Dieckmann et al. (2006)]{Dieckmann06a}
  Dieckmann, M. E., Shukla, P. K. \& Eliasson, B. 2006, Phys. Plasmas 13, 062905

\bibitem[Ellison \& Double (2002)]{Ellison02}
  Ellison, D. C. \& Double, G. P. 2002,Astropar. Physics 18, 213

\bibitem[Krymskii (1977)]{Krymskii77} 
  Krymskii, G. F. 1977, Sov. Phys. Dokl {22}, 6 

\bibitem[Axford, Leer \& Skadron (1977)]{Axford77} 
  Axford, W. I., Leer, E. \& Skadron, G. 1977, Proc. 15th Int. Cosmic Ray Conf., Plovdiv (Budapest: Central  Research Institute for Physics) { 11}, 132 

\bibitem[Bell (1978)]{Bell78} 
  Bell, A. R. 1978, Mon. Not. R. Astron. Soc. {182}, 147

\bibitem[Blandford \& Ostriker (1978)]{Blandford78} 
  R. D. Blandford \& J. Ostriker, Astrophys. J. {\bf 221}, L29 (1978)

\bibitem[Waxman (1997)]{grb_s} 
  E. Waxman, Ap. J. {\bf 485}, L5 (1997)

\bibitem[Bednarz \&  Ostrowski (1998)]{Bednarz98} 
 J. Bednarz \& M. Ostrowski, Phys. Rev. Lett. {\bf 80}, 3911 (1998)

\bibitem[Kirk et al. (2000)]{Kirk00} 
  J. K. Kirk, A. W. Guthmann, Y. A. Gallant, \& A. Achterberg, Phys. Rev. {\bf 542}, 235 (2000) 

\bibitem[Achterberg et al. (2001)]{Achterberg01} 
  A. Achterberg, Y. A. Gallant, J. K. Kirk, \& A. W. Guthmann, Mon. Not. R. Astron. Soc. {\bf 328}, 393 (2001)

\bibitem[Meli \& Quenby (2003)]{Meli03} 
  A. Meli \& J. J. Quenby, Astroparticle Phys., {\bf 19}, 649 (2003)

\bibitem[Ostrowski \& Bednarz( 2002)]{Ostrowski02} 
  M. Ostrowski \& J. Bednarz, Astron. Astrophys. {\bf 394}, 1141 (2002)

\bibitem[{{Keshet} \& {Waxman}(2005)}]{Keshet05}
{Keshet}, U. \& {Waxman}, E. 2005, Physical Review Letters, 94, 111102

\bibitem[{{Keshet} {et~al.}(2004){Keshet}, {Waxman}, \& {Loeb}}]{Keshet04}
{Keshet}, U., {Waxman}, E., \& {Loeb}, A. 2004, Ap. J. 617, 281

\bibitem[{{Drury} {et~al.}(1989)}]{Drury89}
{Drury}, L. O'C., {Markiewicz}, W. J., \& {V{\"o}lk}, H.~J. 1989,
A\&A 225, 179

\bibitem[{Li \& Waxman (2006)}]{Zhuo06}
Li, Z. \& Waxman, E. 2006, arXiv:astro-ph/0603427

\bibitem{CR_data_rev}
  M. Nagano \& A. A. Watson, Rev. Mod. Phys. 72, 689 (2000)

\bibitem[Waxman (2004)]{WNobel04}
  E. Waxman, Proc. Nobel Symposium 129: Neutrino Physics (astro-ph/0502159).

\bibitem[Waxman (2006)]{WPylos06}
  E. Waxman Nucl. Phys. B (Proc. Suppl.) {\bf 151}, 46--53 (2006); astro-ph/0412554.

\bibitem[{{Dieckmann} {et~al}(2006)}]{Dieckmann06}
{Dieckmann}, M. E., {Shukla}, P. K. \& {Drury}, L. O. C. 2006,
Mon. Not. Royal Astron. Soc. 367, 1072

\bibitem[{{Wallace} \& {Epperlein}(1991)}]{Wallace91}
{Wallace}, J. M. \& {Epperlein}, E. M. 1991, Phys. Fluids B, 3,
1579

\bibitem[{{Kato}(2005)}]{Kato05}
Kato, T. N. 2005, Phys. Plasmas 12, 080705

\bibitem[{{Gruzinov}(2001b)}]{Gruzinov01b}
  A. Gruzinov 2001b, arXiv:astro-ph/0111321

\bibitem[{{Medvedev} {et~al.}(2005){Medvedev}, {Fiore}, {Fonseca}, {Silva}, \&
  {Mori}}]{Medvedev05}
{Medvedev}, M.~V., {Fiore}, M., {Fonseca}, R.~A., {Silva}, L.~O., \&
{Mori},
  W.~B. 2005, Ap. J. 618, L75
  
\bibitem[{{Nishikawa} {et~al.}(2003){Nishikawa}, {Hardee}, {Richardson},
  {Preece}, {Sol}, \& {Fishman}}]{Nishikawa03}
{Nishikawa}, K.-I., {Hardee}, P., {Richardson}, G., {Preece}, R.,
{Sol}, H., \&
  {Fishman}, G.~J. 2003, Ap. J. 595, 555

\bibitem[{{Frederiksen} {et~al.}(2004){Frederiksen}, {Hededal}, {Haugb{\o}lle},
  \& {Nordlund}}]{Frederiksen04}
{Frederiksen}, J.~T., {Hededal}, C.~B., {Haugb{\o}lle}, T., \&
{Nordlund},
  {\AA}. 2004, Ap. J. 608, L13

\bibitem[{Spitkovsky(2005)}]{Spitkovsky05}
Spitkovsky, A. 2005, AIP Conf. Proc., 801,
345,~arXiv:astro-ph/0603211

\bibitem[{{Silva} {et~al.}(2003){Silva}, {Fonseca}, {Tonge}, {Dawson}, {Mori},
  \& {Medvedev}}]{Silva03}
{Silva}, L.~O., {Fonseca}, R.~A., {Tonge}, J.~W., {Dawson}, J.~M.,
{Mori},
  W.~B., \& {Medvedev}, M.~V. 2003, Ap. J. 596, L121

\bibitem[{{Jaroschek} {et~al.}(2004){Jaroschek}, {Lesch}, \&
  {Treumann}}]{Jaroschek04}
{Jaroschek}, C.~H., {Lesch}, H., \& {Treumann}, R.~A. 2004, Ap. J. 
616, 1065

\bibitem[{{Romanov} {et~al.}(2004){Romanov}, {Bychenkov}, {Rozmus}, {Capjack},
  \& {Fedosejevs}}]{Romanov04}
{Romanov}, D.~V., {Bychenkov}, V.~Y., {Rozmus}, W., {Capjack},
C.~E., \&
  {Fedosejevs}, R. 2004, Phys. Rev. Lett. 93, 215004

\bibitem[Z\`{e}ldovich \& Raizer (1968)]{zel68}
  Z\`{e}ldovich, Ya. B. \& Raizer, Yu. P. 1968,
  Physics of Shock Waves and High-Temperature Hydrodynamic Phenomena, Chapter XII (New York:Academic).

\bibitem[Kadanoff et al.(1967)]{Kadanoff67} Kadanoff, L.~P., et
al.\ 1967, Reviews of Modern Physics, 39, 395

\end{thebibliography}
\end{document}